\documentclass[a4paper,fleqn,usenatbib]{mnras}

\usepackage{newtxtext,newtxmath}
\usepackage[normalem]{ulem}

\usepackage[T1]{fontenc}
\usepackage{ae,aecompl}

\usepackage{graphicx}	% Including figure files
\usepackage{amsmath}	% Advanced maths commands
\usepackage{amssymb}	% Extra maths symbols
\usepackage{natbib}	% bibliography
\title[Hyades white dwarfs]{A \textit{Gaia}\,DR2 view of white dwarfs in the Hyades }

\author[Salaris \& Bedin]{
M. Salaris$^{1}$\thanks{E-mail: M.Salaris@ljmu.ac.uk}
and L.~R. Bedin$^{2}$
\\
$^{1}$Astrophysics Research Institute, Liverpool John Moores
University,146 Brownlow Hill, Liverpool L3 5RF, UK\\ $^{2}$Istituto
Nazionale di Astrofisica -- Osservatorio Astronomico di Padova, Vicolo
dell'Osservatorio 5, I-35122 Padova, Italy\\ }

\date{Accepted XXX. Received YYY; in original form ZZZ}

\pubyear{2015}

\begin{document}
\label{firstpage}
\pagerange{\pageref{firstpage}--\pageref{lastpage}}
\maketitle

\begin{abstract}
  We have exploited the very precise parallaxes, proper motions and
  photometry of \textit{Gaia}\, Data Release~2 to study white dwarf
  members of the Hyades star cluster.
  \textit{Gaia}\, photometry and parallaxes for the eight DA white
  dwarfs confirmed members have been then used to compute absolute
  magnitudes and colours. These were compared to three independent
  sets of white dwarf evolutionary tracks, to derive cooling times and
  white dwarf (final) masses.  All sets of models provide the same
  mass values, with only small differences in the cooling ages.  The
  precision in the derived masses and cooling ages is typically
  1-3\%. Our derived masses are generally consistent with
  spectroscopic estimates from the literature, whilst cooling ages are
  generally larger.
  The recent estimate of the cluster age from the \textit{Gaia}\, Data
  Release~2 main sequence turn off colour-magnitude-diagram (790~Myr)
  has been employed to derive progenitor (initial) masses.  We find a
  slope of the initial-final mass relation for the Hyades white dwarfs
  (masses between $\sim$0.67 and $\sim 0.84 M_{\odot}$) steeper than
  that derived for the same mass range from \textit{global} estimates
  --averaged over the whole spectrum of white dwarf masses--
  irrespectively of the cooling models adopted.  However, when
  considering the error in this age estimate ($^{+160}_{-100}$ Myr), a
  definitive conclusion on this issue cannot be reached yet.  The
  lower limit of 690~Myr (closer to the \textit{classical} Hyades age
  of 600-650~Myr) would provide a slope of the initial-final mass
  relation closer to the global determinations.
  We also find hints of an intrinsic spread of the cluster
  initial-final mass relation for the cluster.
\end{abstract}


%%%%%%%%%%%%%%%%%%%%%%%%%%%%%%%%%%%%%%%%%%%%%%%%%%

%%%%%%%%%%%%%%%%% BODY OF PAPER %%%%%%%%%%%%%%%%%%

\section{Introduction}

The recent \textit{Gaia} Data Release 2 (DR2) has delivered
high-precision astrometry and three-band photometry (${\rm G, G_{BP},
  G_{RP}}$) of about 1.3 billion sources over the whole sky, with
unprecedented accuracy and homogeneity, of both astrometry and
photometry \citep[]{gaiadr2}.  Indeed, \textit{Gaia}
colour-magnitude-diagrams (CMDs) of the closest open clusters
\citep{gaiaclust} display exquisitely defined sequences in the CMD.
The distance modulus corrected CMD of the classical Hyades cluster,
for example, has typical errors (including the parallax error
contribution) of a few mmag in all three \textit{Gaia} filters, also
along the white dwarf (WD) sequence. Moreover, the distances and proper motions
provided by DR2 allow accurate cluster membership analyses.

Here we focus on the WD sequence hosted by this cluster.  Taking
advantage of the high precision of DR2 parallaxes and photometry, WD
masses and cooling times can be determined very precisely employing
theoretical cooling models. Their initial-final mass relation (IFMR)
can then be established from the knowledge of the cluster age.

The IFMR for low- and intermediate-mass stars is an essential input
for a range of astrophysical problems.  Given an initial stellar mass
on the main sequence (MS), the IFMR provides the expected final
WD mass, hence the total amount of mass lost during the star
evolution.  Not only the location and shape of cooling sequences and
the shape of WD luminosity functions -- employed to age date stellar
populations-- are affected by the IFMR, but also the chemical evolution
histories of stellar populations, as well as their mass-to-light
ratios (the ratio of the mass of evolving stars plus stellar remnants
to the integrated luminosity of the population) and the modelling of
stellar feedback in galaxy formation simulations
\citep[e.g.][]{ak15}. Type Ia supernova rate estimates are affected by
the choice of the IFMR \citep[e.g.][]{greggio} as well.

Theoretical determinations of the IFMR based on stellar evolution
calculations that follow the evolution of stellar models from the
pre-MS to the WD phase are still affected by sizable uncertainties.
This is due to the poorly modelled efficiency of mass-loss for low-
and intermediate-mass stars, and uncertainties in the predicted mass
of CO cores during the asymptotic giant branch (AGB) evolution,
resulting from outstanding uncertainties in the treatment of the
thermal pulse phase, the associated third dredge-up, hot bottom
burning and also the treatment of rotation
\citep[see, e.g.][]{iberen83, domi96, kl14}.  Semi-empirical methods
have been therefore devised to establish the IFMR independently of
theoretical modelling of the AGB phase \citep[see, e.g.,][for recent
  examples]{weide00, fer05, kalirai09, ssw09, williams09, cumm15}.

The \lq{classical\rq} semi empirical technique to estimate the IFMR is
based on WDs in star clusters, and works as follows.  Spectroscopic
analyses provide the WD surface gravity $g$ and $T_{\rm eff}$, and
for a given $g-T_{\rm eff}$ pair, grids of theoretical WD models
provide the mass $M_{\rm f}$ and cooling age $t_{\rm cool}$ of
the WD.  Theoretical isochrone fits to the MS turn-off luminosity in
the cluster CMD provide the cluster age
$t_{\rm cl}$. Finally, the difference $t_{\rm cl} - t_{\rm
  cool}$ is equal to the lifetime $t_{\rm prog}$ of the WD
progenitor from the MS until the start of the WD cooling. Making use
of mass-lifetime relationships from theoretical stellar evolution
models, the initial progenitor mass $M_{\rm i}$ is immediately
obtained from $t_{\rm prog}$ (the uncertain AGB and post-AGB
lifetimes can be neglected, because they are negligible compared to the
duration of the previous evolutionary phases).

Very recently, the \textit{Gaia} DR2 CMD of
6400 bright WDs within a distance of 100~pc has been employed by \citet{badry18} to place strong 
constraints on the IFMR, especially for $M_{\rm i} < 4 M_{\odot}$.  
These authors assumed an age distribution for the WDs, assessed the completeness of their sample, and 
determined the WD masses from fits of cooling tracks to the observed CMD position of each individual objects.
Their derived IFMR is broadly consistent with current star cluster studies.

Here we focus on the \textit{Gaia} DR2 CMD of the Hyades WDs. Previous
analyses, in particular \citet{cumm15}, have shown that studies of
clusters in the age range of the Hyades provide an IFMR for $M_{\rm i}$
between $\sim 2.5$ and $\sim 4.0 M_{\odot}$, that displays a
slope much steeper than what obtained fitting an average relationship
over a much broader mass range. Also, there are hints of maybe an intrinsic dispersion of the
IFMR in this $M_{\rm i}$ range \citep[e.g.][]{ssw09}. These are clearly
important issues, that we are going to revisit taking advantage of the
new \textit{Gaia} data.  DR2 provide very accurate photometry and
parallaxes (fractional errors in the order of $10^{-3}$) for the
Hyades WDs, allowing us to determine the IFMR by deriving $M_{\rm f}$
and $t_{\rm cool}$ from fits of theoretical cooling sequences to the
WD \textit{Gaia} CMD \citep[as done by][for field WDs]{badry18}.

Moreover, \textit{Gaia} DR2 data will allow us to determine whether
the sample of new Hyades WD candidates discussed in \citet{tsr12} 
contains truly Hyades stars.

The plan of the paper is as
follows. Section~\ref{data} describes the Hyades WD sample and the
membership of \citet{tsr12} candidates, whilst section~\ref{analysis}
describes our derivation of the IFMR and associated errors.
Section~\ref{comparisons} compares our inferred IFMR with previous
determinations, and conclusions follow in Sect.~\ref{conclusions}.

\section{Data}
\label{data}

The 518 members for the Hyades cluster adopted for this work,
are those defined and released by \citet[Table\,A.1]{gaiadr2}. 

For such a close-by cluster the projection effects of the cluster mean 
radial velocity on individual members can be as large as 41 mas/yr,
due to the large spread over the sky (over 30 degrees) and in
parallax space. The projection of the tangential motion is also
sizable for the Hyades, amounting to about 5~mas/yr.  There is also a
scaling effect, do to the non-negligible difference in distance of the 
members (the cluster radius of $\sim$15 pc is comparable to its
distance, $\sim$50 pc).

The method applied to determine the cluster membership is detailed in  
\citet{gaiadr1}, specifically for the Hyades cluster. 
In essence, in the \lq{combined astrometric solution\rq} the observed
parallaxes and proper motions are compared with predicted ones,
calculated with the current assumed parallax and space motion of
the cluster centre, and with position of the star relative to the
projected cluster centre.  The improved knowledge of the consistency
of the motions of stars within the searching volume allows to better
define the sample of members, and in turn, to improve the motion and
parallax of the cluster centre, iteratively.

\begin{figure}
	\includegraphics[width=\columnwidth]{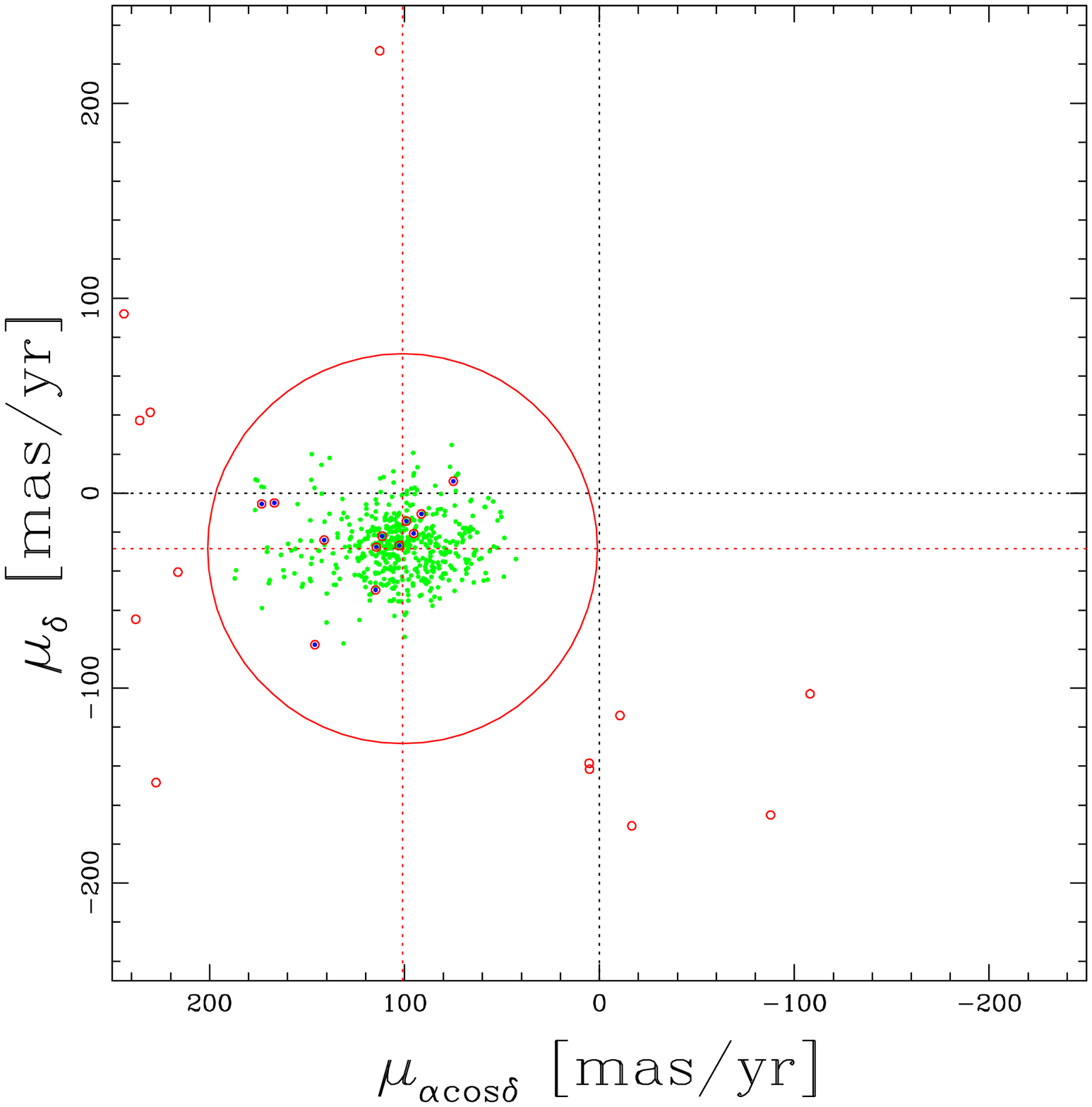}
        \caption{
          Proper motion diagram for the Hyades. 
          Green dots show the Hyades members as defined by \citet{gaiaclust}. Dotted lines mark the zero
          motion (in black), and the motion of the cluster's centre
          (in red) as defined in \citet{gaiaclust}. Red dots display the 25 WD candidate members in
          Table\,2 of \citet{tsr12}.  A red circle with a
          radius of 100\,mas\,yr$^{-1}$ arbitrarily sets a
          proper-motion membership criterion (see text).}
    \label{fig:selection}
\end{figure}

Figure~\ref{fig:selection} displays the proper-motion diagram for the
Hyades members as defined by \citet{gaiaclust}.  In this diagram, we also show the \textit{Gaia}\,DR2 proper
motions for the 25 WD cluster member candidates defined in Table\,2 of
\citet{tsr12}; only 9 out of these 25 objects are defined as
members by the \textit{Gaia} team.
Given the large motion of the cluster, and the fact that proper
motions are better constrained than parallaxes, we relax the
membership selections of \citet{gaiaclust} and 
arbitrarily use only proper motions as membership criterion for the 25
objects in \citet{tsr12} Table\,2. We set the proper motion threshold for
membership as large as the largest motions in the Gaia member sample,
approximated to the next round number.  We generously set this limit to
100\,mas\,yr$^{-1}$ (red circle). Even with such a \textit{relaxed} limit,
based solely on proper motions, 13 objects are clearly excluded as
members.
Of the 12 WD member candidates that survive this mild proper
motion selection, 9 were already in the \textit{Gaia} selection,  
and only 3 are potential additional WDs candidates. 
Seven of the 9 WDs surviving the \textit{Gaia} selection --also included in \citet{tsr12} paper-- are
\lq{classical\rq} Hyades WDs as defined by \citet{tsr12} and
previously used in IFMR determinations \citep[see, e.g.,][]{fer05,
  ssw09}.  Two other objects (HG~7-85, and GD~52) belong to the
\lq{new candidates\rq} listed by \citet{tsr12}.

\begin{figure}
	\includegraphics[width=\columnwidth]{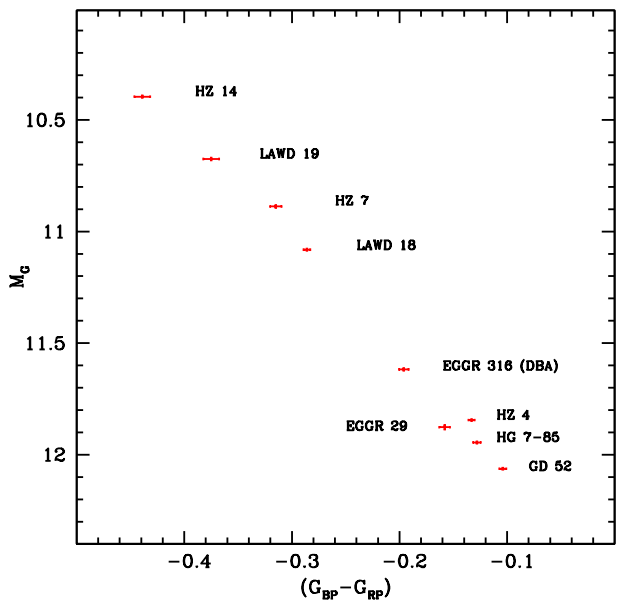}
        \caption{\textit{Gaia} DR2 CMD --distance modulus corrected--
          of the sample of 9 known Hyades WD members. Error bars
          include the DR2 quoted photometric errors and the
          contribution from the parallax error. 
}
    \label{fig:sample}
\end{figure}

The \textit{Gaia} CMD of these nine WDs is shown in
Fig.~\ref{fig:sample}. They are all DA objects as reported by the
Montreal White Dwarf Database \citep{mwdd}, but for 
EGGR~316, that is
a DBA WD with mixed H/He atmosphere \citep{berge11}. Due to the lack
of extended grids of cooling models and bolometric corrections for DBA
objects, we have not included this WD in our analysis.  Individual
parallaxes\footnote{We did not apply a correction for the
  $\sim$$-$0.03~mas offset of DR2 parallaxes \citep{gaiaoffset}, for
  its effect on the distance of the Hyades' stars is absolutely
  negligible.}, their fractional errors, absolute magnitudes in the
\textit{Gaia} G filter, as well as the $\rm {(G_{BP}-G_{RP})}$ colours
and associated $1\sigma$ errors (taking into account also the errors
on the parallax) are reported in Table~\ref{tab:dataWD}.
The accuracy of photometry and parallax measurements provide absolute
magnitudes and colour uncertainties well below 0.01~mag.

\begin{table*}
	\centering
	\caption{Data about the 8 DA Hyades WDs shown in Fig.~\ref{fig:sample}. We display, from left to right, 
          WD name,
          \texttt{Identifier:\,Gaia\,DR2},
          parallax (in mas), parallax fractional error, absolute ${\rm  G}$ magnitude with error (including the
          contribution from the parallax error), colour with associated error, logarithm of the cooling time (in years)
          obtained with $(a)$ the \citet{bastiwd} models and error, logarithm of the cooling time 
          obtained with $(b)$ the \citet{renedo10} models and error, mass (in Solar units) and associated error.}
	\label{tab:dataWD}
	\begin{tabular}{rrccccccc} 
		\hline
		Name & \texttt{Identifier:\,Gaia\,DR2} & $\pi$ & $\sigma_{\pi}/\pi$ & ${\rm M_G}\pm \sigma$ & ${\rm (G_{BP}-G_{RP})\pm\sigma}$& ${\rm log}(t^a_{\rm cool})\pm\sigma$& ${\rm log}(t^b_{\rm cool})\pm\sigma$ & $M_{\rm f}\pm\sigma$\\
               (1) & (2) & (3) & (4) & (5) & (6) & (7) & (8) & (9)\\
		\hline
              HZ~14  & \texttt{3294248609046258048} & 20.25& 0.0025 &10.40$\pm$0.005 &$-$0.439$\pm$0.007 &7.430$\pm$0.010 &7.370$\pm$0.010&0.71$\pm$0.02\\
              LAWD~19& \texttt{3313714023603261568} & 20.89& 0.0027 &10.68$\pm$0.006 &$-$0.375$\pm$0.007 &7.730$\pm$0.010 &7.710$\pm$0.010&0.69$\pm$0.02\\
              HZ~7   & \texttt{3306722607119077120} & 21.14& 0.0029 &10.89$\pm$0.006 &$-$0.315$\pm$0.005 &7.940$\pm$0.010 &7.940$\pm$0.010&0.67$\pm$0.02\\
              LAWD~18& \texttt{3313606340183243136} & 22.23& 0.0023 &11.08$\pm$0.005 &$-$0.286$\pm$0.003 &8.083$\pm$0.007 &8.083$\pm$0.007&0.69$\pm$0.01\\
              HZ~4   & \texttt{3302846072717868416} & 28.59& 0.0019 &11.84$\pm$0.004 &$-$0.133$\pm$0.003 &8.543$\pm$0.006 &8.573$\pm$0.006&0.79$\pm$0.01\\
              EGGR~29& \texttt{45980377978968064}   & 19.94& 0.0047 &11.88$\pm$0.010 &$-$0.158$\pm$0.005 &8.545$\pm$0.007 &8.575$\pm$0.007&0.83$\pm$0.01\\
              HG~7-85& \texttt{3306722607119077120} & 24.05& 0.0022 &11.94$\pm$0.005 &$-$0.128$\pm$0.004 &8.595$\pm$0.007 &8.625$\pm$0.007&0.82$\pm$0.01\\
              GD~52  & \texttt{218783542413339648}  & 23.56& 0.0019 &12.06$\pm$0.004 &$-$0.104$\pm$0.003 &8.670$\pm$0.007 &8.700$\pm$0.007&0.84$\pm$0.01\\
		\hline
	\end{tabular}
\end{table*}

\begin{figure}
	\includegraphics[width=\columnwidth]{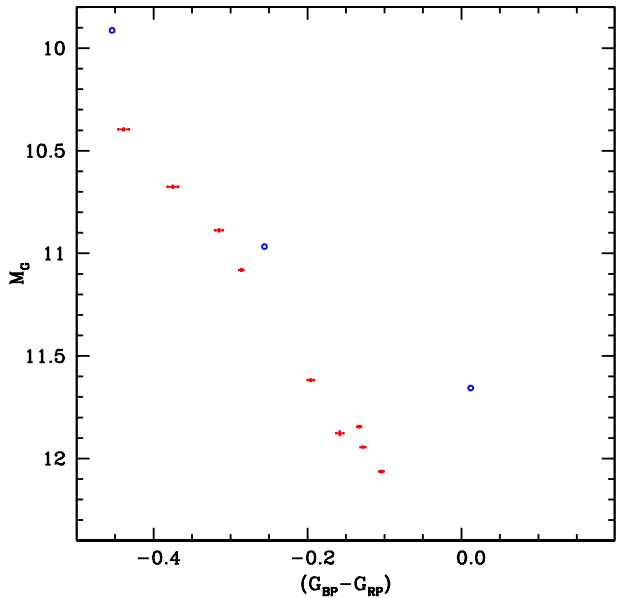}
        \caption{As Fig.~\ref{fig:sample} but excluding the DBA object
          EGGR~316, and including GD~38, GD~43 and LP~475-249 (open
          circles without error bars -- see text for details).}
    \label{fig:excluded}
\end{figure}

The remaining three WDs that survive our proper motion selection \citep[GD~38, GD~43 and LP~475-249, all DA according to][database]{mwdd}
have {\it suspiciously} low parallaxes, of the order of 7-10~mas,
compared to 20-30~mas for the nine WDs of Fig.\,\ref{fig:sample}.
Figure\,\ref{fig:excluded} shows the \textit{Gaia} CMD of these three
objects, together with the other WDs.
They seem to occupy a redder sequence compared to the nine confirmed member WDs. 
Notice that \citet{tsr12} at the end of their analysis considered
these objects \lq{non-members\rq}.

The quality indicators available in
  \textit{Gaia}\,DR2\footnote{\texttt{https://gea.esac.esa.int/archive/documentation/GDR2/}}
  have been inspected for these three WDs and the objects in
  Table~\ref{tab:dataWD}. 
More specifically, we have considered: 
\texttt{visibili\-ty\_per\-io\-ds\_used,
astro\-metric\_matched\_observations,
astro\-metric\_gof\_al,
astro\-metric\_excess\_noise,
astro\-metric\_n\_good\_obs\_al,
astro\-metric\_n\_bad\_obs\_al}. 
None of these quality indicators turned out to be 
significantly worse than for the 8~WDs in Table~\ref{tab:dataWD}, 
suggesting that the smaller parallaxes for these three objects are equally reliable.

The parallax errors for the WDs in Table~\ref{tab:dataWD} span the
range 0.052-0.062~mas (with the exception of EGGR\,29, with a parallax
error of 0.092\,mas), whilst for these three objects the errors are
typically larger, spanning the range 0.075-0.115~mas.

Concerning the photometry, the fluxes appear to have marginally
larger errors, particularly in G$_{\rm RP}$, but there is no evidence of differences in the
\texttt{phot\_bp\_rp\_excess\_factor} parameter, that remains within
the values observed for the other WDs.
Finally, we note that all the WDs in Table~\ref{tab:dataWD} 
and these three objects passed the tests for well-measured objects defined in Eqs.~(C.1) and (C.2) of 
\citet{lindegr18}.

One could then suspect that these three WDs might suffer from
systematic errors in their parallax (and radial velocities) due for
example to the presence of close binary companions.  Future
\textit{Gaia} data releases will very likely clarify the situation
regarding these objects.  We will not include them in the analysis
that follows.

%%%
%%%

\section{Analysis}
\label{analysis}

\begin{figure}
	\includegraphics[width=\columnwidth]{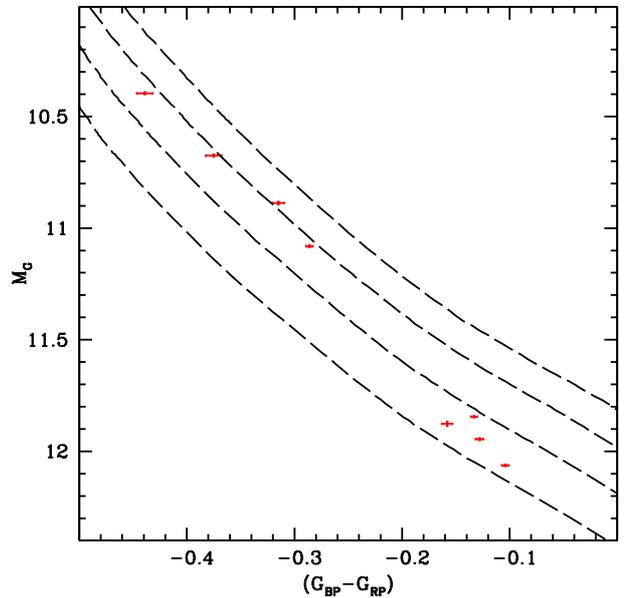}
        \caption{CMD of the eight DA WDs of Table~\ref{tab:dataWD}
          employed in our IFMR analysis, together with the
          \citet{bastiwd} cooling tracks for masses equal to 0.61,
          0.68, 0.77 and 0.87~$M_\odot$ (see text for details).}
    \label{fig:cooling}
\end{figure}

Figure~\ref{fig:cooling} displays the CMD of our final sample of eight
DA WDs together with our reference DA cooling tracks for masses equal to
0.61, 0.68, 0.77 and 0.87~$M_\odot$, from \citet{bastiwd}. The cooling
tracks are calculated for CO cores \citep[see][for details about the
  CO stratification]{bastiwd} and thick H layers ($10^{-4}M_{\rm WD}$,
on top of a $10^{-2}M_{\rm WD}$ He layer). Bolometric corrections to
the {\textit Gaia} DR2 system have been kindly provided by P. Bergeron
\citep[private communication, see][]{hb06, tbg11}.

Interpolation amongst the cooling tracks to match ${\rm M_G}$ and
${\rm (G_{BP}-G_{RP})}$ of each individual WD \citep[we assumed zero reddening for the cluster, see e.g.][]{t06} provides
straightforwardly mass and cooling age, also reported in
Table~\ref{tab:dataWD}. To estimate the associated errors, we have generated for each object  
one thousand synthetic ${\rm M_G}$ and 
${\rm (G_{BP}-G_{RP})}$ pairs, with Gaussian distributions (assumed to be independent) centred around the measured values, 
and 1$\sigma$ widths equal to the errors on these quantities reported in Table~\ref{tab:dataWD}. 
Mass and cooling times for each synthetic sample were then determined from the WD tracks, and the 68\% confidence limits 
calculated.

These formal errors --determined by the error
bars on absolute magnitudes and colours-- are small \citep[smaller
  than in previous error estimates, see e.g.,][]{tsr12}, and equal to
0.01-0.02 $M_{\odot}$ in the derived masses, and $\sim$0.01 or less in
log($t_{\rm cool}$).  We notice that all WDs in this sample have evolved beyond the
luminosity range where neutrino energy losses dominate (log$L/L_{\odot}$ above $\sim -1$~dex), but have not
yet started crystallization, and none of them has a cooling age very close
to the cluster age (see below).

Although formal errors are small, systematics due to uncertainties in
WD modelling might add non-negligible systematic components. To
estimate these effects, we have considered the independent DA WD
calculations by \citet{fbb01} and \citet{renedo10}, that employ some
different physics inputs compared to \citet{bastiwd}. We applied to
these cooling tracks the same bolometric corrections as for our
reference WD models.  The models by \citet{fbb01} that we employed are
for CO cores \citep[although with a different stratification compared
  to ][]{bastiwd} and the same envelope composition and
thickness of \citet{bastiwd}.  We obtain for our sample
indistinguishable values of both cooling ages and masses, compared to
that
reported in Table~\ref{tab:dataWD}.

The calculations by \citet{renedo10} are fully evolutionary, in the
sense that the WD tracks come from the complete evolution of a MS
progenitor (hence with assumptions about the mass loss along the AGB
phase) with a given initial metallicity. We used the models for
initial metal mass fraction about half Solar, the highest value
available in \citet{renedo10} calculations, but lower than the Hyades
spectroscopic measurement of [Fe/H]$\sim$0.10-0.15 \citep[see
  e.g.][]{tj05, df16}. We have however verified by comparing the half
Solar cooling tracks with the ones at lower metallicity (a factor of
ten lower) from the same \citet{renedo10} paper, that the cooling
times and CMD location in the magnitude range relevant to our
analysis, are practically identical. This suggests that, at least for
the magnitude range of the Hyades WDs, the initial metallicity of these WD model progenitors 
does not affect the main properties of the cooling tracks.
The \citet{renedo10} WD models have a CO
stratification different from \citet{bastiwd} and \citet{fbb01}, and
also a mass thickness of of the H and He layers that varies with WD
mass\footnote{In addition, the H and He profiles are not step function like in \citet{bastiwd} and \citet{fbb01} but
have been shaped by atomic diffusion during the early WD phases}. Values (decreasing with increasing WD mass in the range between
0.53 and 0.88 $M_{\odot}$) go from $10^{-3.6}M_{\rm WD}$ to
$10^{-4.9}M_{\rm WD}$ for the H layers, and from $10^{-1.6}M_{\rm WD}$
to $10^{-2.9}M_{\rm WD}$ for the He layers, and are overall not too
different from \citet{bastiwd} and \citet{fbb01} models.

\begin{table}
	\centering
	\caption{Initial masses estimated for the 8 DA WDs of Fig.~\ref{fig:cooling}. From left to right
          we display the WD name, the initial mass (in Solar masses) and the asymmetric error bars estimated from
          the \citet{bastiwd} cooling times, and the \citet{renedo10} ones, respectively.}
	\label{tab:ifmr}
	\begin{tabular}{rcccccc} 
		\hline
		Name & ${\rm M^a_i}$ & ${\Delta^{-}}$ & ${\Delta^{+}}$ & ${\rm M^b_i}$& ${\rm \Delta^{-}}$& ${\rm \Delta^{+}}$\\
		\hline
              HZ~14   & 2.53 & 0.16 &  0.12 & 2.52 & 0.16 &0.12\\
              LAWD~19 & 2.55 & 0.16 &  0.12 & 2.55 & 0.16 &0.12\\
              HZ~7    & 2.60 & 0.18 &  0.14 & 2.60 & 0.18 &0.14\\
              LAWD~18 & 2.64 & 0.19 &  0.15 & 2.64 & 0.19 &0.15\\
              HZ~4    & 3.05 & 0.34 &  0.29 & 3.11 & 0.36 &0.31\\
              EGGR~29 & 3.05 & 0.34 &  0.29 & 3.11 & 0.36 &0.31\\
              HG~7-85 & 3.16 & 0.39 &  0.34 & 3.25 & 0.42 &0.38\\
              GD~52   & 3.41 & 0.50 &  0.47 & 3.55 & 0.57 &0.56\\
		\hline
	\end{tabular}
\end{table}

\begin{figure}
	\includegraphics[width=\columnwidth]{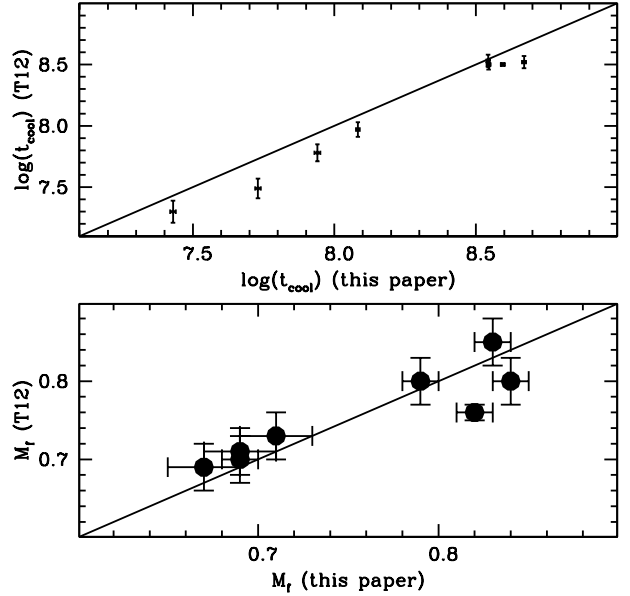}
        \caption{Comparison of $t_{\rm cool}$ (upper panel) and
%         $M_{\rm f}$ (lower panel) between our results (columns of
%         8-9 of Table~\ref{tab:dataWD} and columns 2-4 of
          $M_{\rm f}$ (lower panel) between our results (columns (7) and (9) of
          Table~\ref{tab:dataWD}) and the corresponding values from
          \citet{tsr12}.}
    \label{fig:comp}
\end{figure}

With this additional grid of WD models we have redetermined the Hyades WD
masses and cooling ages. Masses are unchanged compared to the results
with \citet{bastiwd} and \citet{fbb01} models, whereas cooling ages
are slightly different, i.e. longer at higher WD masses, and unchanged
or slightly shorter at the lower masses. This second set of cooling
ages is also reported in Table~\ref{tab:dataWD}.  Finally, we made use
of the analysis by \citet{ssw09} to assess the effect of decreasing
the mass thickness of the hydrogen layers of the models by two orders
of magnitude, below the standard \lq{thick\rq} layers value of
$10^{-4}M_{\rm WD}$. For the magnitude and colour range of our sample
of Hyades WDs the effect on the derived masses and cooling times is
smaller than the errors reported in Table~\ref{tab:dataWD}.
 
Figure~\ref{fig:comp} compares our determination of WD masses $M_{\rm
  f}$ and cooling times \citep[we display only the result obtained
  with our reference models, the comparison is very similar when
  considering the cooling times obtained with][models]{renedo10} with
the corresponding values listed by \citet{tsr12}.  The $M_{\rm f}$
values in \citet{tsr12} are taken from the literature, whilst cooling
times come from fits of the \citet{fbb01} DA models to literature
values of $g-T_{\rm eff}$ pairs for each object. The WD masses are
pretty much in agreement within the errors -- apart from HG~7-85, whose
mass is equal 0.76$\pm0.01~M_{\odot}$ in \citet{tsr12},
whereas we find 0.82$\pm 0.01 M_{\odot}$-- and
log($t_{\rm cool})$ values are typically larger than \citet{tsr12}.

Having determined precise WD masses and cooling ages from the CMD, we
need a cluster age from the MS turn off. \citet{gaiaclust} provide an 
age estimated from \textit{Gaia} DR2 MS photometry and \citet{parsec} 
isochrones for [Fe/H]=0.13, transformed to the \textit{Gaia} DR2
photometric system. This is equal to
log($t_{\rm  cl}$)=8.90$^{+0.08}_{-0.06}$ ($t_{\rm cl}$ in years). 
Using this value (and error bar), we have determined $M_{\rm i}$ for our WD
sample considering the two sets of $t_{\rm cool}$ values reported in
Table~\ref{tab:dataWD}, and --consistently with the cluster age
estimate-- the initial mass-lifetime values from \citet{parsec}
evolutionary tracks.

The two sets of $M_{\rm i}$ values we have obtained are shown in
Table~\ref{tab:ifmr}. The error on $M_{\rm i}$ is largely dominated by
the error bar on the cluster age, hence it is essentially a systematic
error on the WD final mass, because increasing or decreasing the
cluster age according to its error bar does systematically decrease or
increase, respectively, the values of $M_{\rm i}$ for all WDs of any $M_{\rm f}$.

\begin{figure}
	\includegraphics[width=\columnwidth]{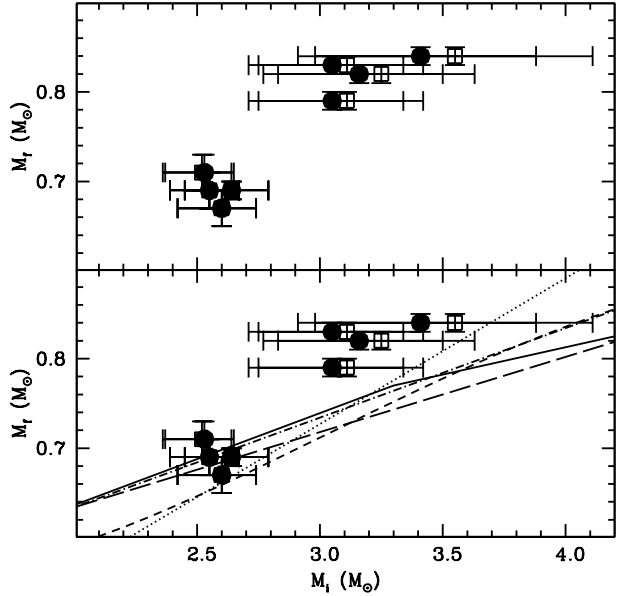}
        \caption{IFMRs we have obtained using \citet{bastiwd} cooling times
          (filled circles) and \citet{renedo10} cooling times
          (open squares).  The two sets of estimates overlap at the lower masses.  The
          lower panel displays also the independent results by
          \citet{fer05} as dot-dashed (linear IFMR) and short-dashed (polynomial IFMR) lines, 
          \citet{ssw09} as a dashed line (their linear IFMR), \citet{badry18} as a solid line, and
          \citet{cumm15} as a dotted line.}
    \label{fig:ifmr}
\end{figure}

\section{Comparisons with previous results}
\label{comparisons}

Figure~\ref{fig:ifmr} shows the IFMRs from the data in
Table~\ref{tab:dataWD} and \ref{tab:ifmr}. Errors in $M_{\rm i}$ range
between $\sim$0.15 and $\sim$0.6 $M_{\odot}$, increasing with
increasing $M_{\rm i}$\footnote{
  The essentially constant error on the
  progenitor ages --dominated by the error on the
  cluster age-- causes a larger error on the initial mass for larger
  $M_{\rm i}$ values, because of increasingly shorter lifetimes with
  increasing initial mass.}.
The effect on $M_{\rm i}$ that arises from the variation of cooling
times caused by different WD models is much smaller than the error bar
due to the error on the cluster age.

The first interesting consideration is the hint of a small spread in
$M_{\rm f}$ at fixed $M_{\rm i}$, at least for the pair HZ~4 and
EGGR~29. Their mass difference is much larger than the associated
errors, whilst their $M_{\rm i}$ is virtually the same.

\begin{figure}
	\includegraphics[width=\columnwidth]{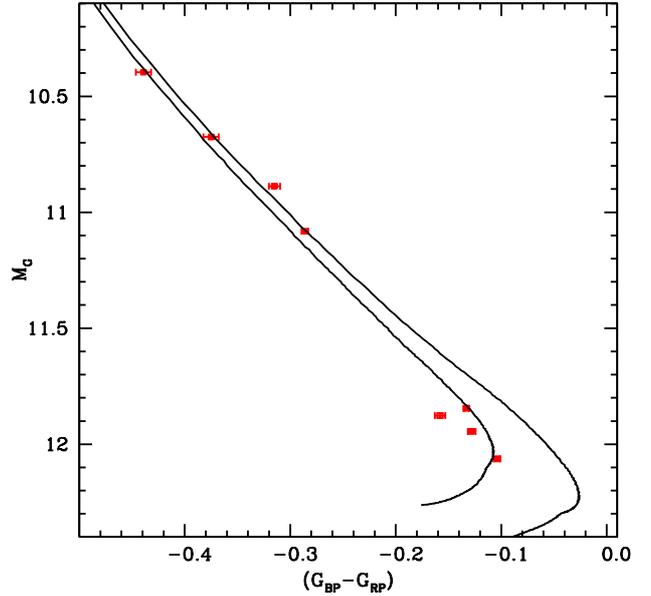}
        \caption{CMD of the eight DA WDs of Table~\ref{tab:dataWD}, compared to a 600~Myr and a 800~Myr WD isochrone
          calculated employing the \citet{ssw09} IFMR (see text for details).}
    \label{fig:iso}
\end{figure}

Regarding the slope of the IFMR, Fig.~\ref{fig:ifmr} displays 
\citet{fer05} -- both linear and polynomial IFMRs--  
\citet{ssw09}\footnote{We display their linear fit to the global IFMR. 
Their two-slope fit gives a very similar IFMR in the displayed mass range} and \citet{badry18} global determinations 
of the IFMR, plus the \citet{cumm15} IFMR determined (from Praesepe, Hyades and NGC~2099) 
for the $M_{\rm i}$ range between $\sim$2.5 and $\sim 4~M_{\odot}$. 
 
If at first we neglect the error bars on $M_{\rm i}$ values, the slopes from the global 
determinations of the IFMR are clearly shallower than our Hyades one, 
confirming the results by \citet{cumm15}. To give an idea of the differences, \citet{fer05} 
and \citet{ssw09} linear IFMRs both give slopes $\Delta M_{\rm f}/\Delta M_{\rm i} \sim 0.10$, 
to be compared to our slope 
$\Delta M_{\rm f}/\Delta M_{\rm i}=0.20$ (considering cooling times from our reference WD models).

The slope derived by \citet{cumm15} (apart from a vertical zero-point shift of their IFMR)
is closer to our results, for they obtain $\Delta M_{\rm f}/\Delta M_{\rm i}=0.163\pm0.022$.
The polynomial IFMR by \citet{fer05} is steeper than the linear one in this mass range, but still shallower than 
 \citet{cumm15}.
These conclusions hold in case of
using both sets of WD cooling times in Table~\ref{tab:dataWD}.

It is however important to consider the role of the uncertainty in the cluster
age. The error bar on $M_{\rm i}$ is essentially systematic, and
determined by the large error on \citet{gaiaclust} Hyades age ($t_{\rm  cl}\sim 790^{+160}_{-100}$~yr).
Given that a fixed variation of
$t_{\rm cl}$ causes larger changes of the initial mass with increasing
$M_{\rm i}$, the slope of the IFMR will depend on the exact value of
$t_{\rm cl}$.  If we consider the lower limit of \citet{gaiaclust}
$t_{\rm cl}$ -- hence all $M_{\rm i}$ values at the upper limit of their 
individual error bars-- the slope of the Hyades IFMR gets
shallower, equal to $\Delta M_{\rm f}/\Delta M_{\rm i}=0.14$,  
hence closer to the slope of the global estimates and marginally lower than \citet{cumm15} result. 
This lower age limit is actually more consistent with the \textit{classical} Hyades age
of $\sim$600-650~Myr \citep[e.g.][]{perr98}. 

Figure~\ref{fig:iso} shows the impact of the IFMR on the WD isochrones
for the Hyades. We have displayed isochrones for 800~Myr (the age
derived from the \textit{Gaia} DR2 data) and 600~Myr (the more
classical Hyades age) respectively, computed using the \citet{ssw09}
IFMR --as representative of the global relationships determined
considering the full range of WD progenitor masses-- the WD tracks by \citet{bastiwd},
and \citet{parsec} progenitor lifetimes.  For a cluster age
of 800~Myr, the isochrone is clearly offset from the data at the
fainter magnitudes. This is consistent with Fig.~\ref{fig:ifmr},
whereby the \citet{ssw09} IFMR predicts WD masses roughly consistent
with the Hyades ones for the lower mass (brighter) WDs, but predicts
too low final masses for the more massive (fainter) cluster WDs.  An
age of 600~Myr gives indeed a better fit to the observed sequence
because, as discussed above, for such an age we would derive a cluster
IFMR in much better agreement with the global estimates.

Clearly, a reduction of the error on the Hyades age determined from the \textit{Gaia} CMD is required,
to reduce the uncertainty on the slope of the cluster IFMR.
We just recall here that other recent estimates of the Hyades age
--not based on DR2 data-- have provided 650$\pm$70~Myr from the
lithium depletion boundary technique applied to cluster brown dwarfs
\citep{ldb18}, 750$\pm$100~Myr from a Bayesian fit of rotating stellar
models to the $Hipparcos$ CMD of the cluster \citep{bh15}, and
700$\pm$100~Myr from non-rotating isochrone fitting to an optical CMD
using the mean distance modulus from \textit{Gaia} Data Release~1 
\citep{bastinew}.

We close this section noticing that at the time of the submission of
our work, \citet{si18} preprint has appeared. These authors have
applied a sophisticated Bayesian technique to fit the whole CMD (their
data are not from \textit{Gaia} DR2) of a sample of clusters
--including the Hyades-- with theoretical isochrones and WD
evolutionary tracks, to determine distances, ages, [Fe/H], WD masses,
cooling ages and the IFMR. They also combined multiple star clusters
into a hierarchical model to redetermine the IFMR, correcting the
cluster-specific analysis by \textit{borrowing strength} from other
clusters.  For the Hyades (in this case these authors employ distances
from Hipparcos), cluster-specific and hierarchical estimates provide
similar (within the associated error bars) IFMRs. Their Hyades WD
sample comprises 6 DA objects, all included in our analysis (HZ~14,
LAWD~19, HZ~7, LAWD~18, HZ~4, EGGR~29).  Their $M_{\rm f}$ estimates
are systematically lower than ours, differences ranging from
$\sim$0.02 to $\sim$0.08~$M_{\odot}$, while initial masses $M_{\rm i}$
are instead larger, by amounts between $\sim$0.2 and
$\sim$0.7~$M_{\odot}$.  The best-fit age derived for the cluster
is equal to $\sim$600~Myr, and this should explain at least
qualitatively their larger values of $M_{\rm i}$ compared to our
analysis.  
The resulting slopes for the cluster IFMR are equal to $\Delta
M_{\rm f}/\Delta M_{\rm i}=$0.20$\pm$0.05 (cluster-specific) and
0.14$\pm$0.06 (hierarchical). 

Application of their 
technique to \textit{Gaia} DR2 data would be welcome, to investigate whether a 
more precise cluster age can be determined, its consistency with the determination by \citet{gaiaclust}, 
and also to test the consistency with our WD masses and cooling times.

\section{Conclusions}
\label{conclusions}

We employed the \textit{Gaia}\,DR2 sample of \textit{bona-fide}
Hyades member stars, and selected among those WD stars.
Seven out of a total of nine \textit{Gaia}\,DR2 WD members 
are \textit{classical Hyades WDs}, and two additional ones are listed
by \citet{tsr12} as \textit{new candidate} members.
Eight objects are of DA spectral 
type, for which we determined masses and cooling times.  

Three more Hyades candidates in \citet{tsr12} list do survive the
proper motion membership analysis, but have systematically lower
parallaxes compared to the other nine objects.  We suspect that they
might suffer from some undisclosed systematic error, such as the
presence of binary companions that may affect their estimated
parallaxes (and in turn their absolute magnitudes).  We have discarded
these three objects in our analysis, however, they might deserve follow-up
investigations.

The accuracy of the \textit{Gaia} parallaxes (errors of the order of 0.1\%)
and photometry (errors of the order order of mmag) allow to
determine precise masses (errors of 1-3\%) and cooling
times (errors of 1-2\%) for these 8 DA WDs, considering also the
effect of varying the set of cooling models and the thickness of the
atmospheric layers on the associated error bars.  An IFMR for the
Hyades WDs has been then determined by assuming the cluster MS turn
off age ($\sim$800~Myr) recently determined also from the \textit{Gaia} DR2 cluster
CMD \citep{gaiaclust}.  Assuming this turn off age, we find 
that in the $M_{\rm i}$ range between $\sim$2.5 and $\sim 4 M_{\odot}$,  
the cluster IFMR is steeper than average global IFMRs independently  
estimated considering the full range of WD masses and
progenitors. 
The error on this \textit{Gaia} DR2 age estimate ($\sim^{+160}_{-100}$
Myr, that translates into a systematic error on the IFMR) is however
large enough to induce a non-negligible variation of the derived IFMR
slope.  A lower limit of 690~Myr for the cluster age (closer to the
\textit{classical} Hyades age of 600-650~Myr) would provide a slope 
in much closer agreement to global determinations.
Recent independent determinations of the cluster age do not help
narrowing down the turn off age estimate, that remains the dominant 
source of uncertainty in the determination of the cluster IFMR.

\section*{Acknowledgments}

We warmly thank Antonella Vallenari and Diego Bossini for useful
information concerning \textit{Gaia} and the \textit{Gaia}\,DR2 catalog of
Hyades released in their \citet{gaiaclust} paper. We are deeply
indebted to Pierre Bergeron who kindly provided us with bolometric
corrections to the \textit{Gaia}\,DR2 system for the WD cooling
tracks.
We are grateful to our referee, Nigel Hambly, for his
  careful reading and comments, that contributed to improve
  our manuscript.
This work presents results from the European Space Agency (ESA) space
mission \textit{Gaia}.  Gaia data are being processed by the
\textit{Gaia} Data Processing and Analysis Consortium (DPAC).  Funding
for the DPAC is provided by national institutions, in particular the
institutions participating in the \textit{Gaia} MultiLateral Agreement
(MLA). The \textit{Gaia} mission website is
https://www.cosmos.esa.int/gaia.  The \textit{Gaia} archive website is
https://archives.esac.esa.int/gaia.
%%%%%%%%%%%%%%%%%%%%%%%%%%%%%%%%%%%%%%%%%%%%%%%%%%



\bibliographystyle{mnras}
\bibliography{Hyades_MiMf} % if your bibtex file is called example.bib

%\begin{thebibliography}{99}
%\end{thebibliography}

%%%%%%%%%%%%%%%%%%%%%%%%%%%%%%%%%%%%%%%%%%%%%%%%%%

%%%%%%%%%%%%%%%%% APPENDICES %%%%%%%%%%%%%%%%%%%%%

%\appendix



%%%%%%%%%%%%%%%%%%%%%%%%%%%%%%%%%%%%%%%%%%%%%%%%%%

% Don't change these lines
\bsp	% typesetting comment
\label{lastpage}
\end{document}